\documentclass[12pt]{article}

\usepackage{newtxtext,newtxmath}
\usepackage[T1]{fontenc}

\usepackage{graphicx}
\usepackage{float}

\usepackage[letterpaper,margin=1in]{geometry}

\renewenvironment{abstract}
	{\quotation}
	{\endquotation}

\date{}

\makeatletter
\renewcommand{\fnum@figure}{\textbf{Figure \thefigure}}
\renewcommand{\fnum@table}{\textbf{Table \thetable}}
\makeatother

\usepackage{scicite}

\usepackage{url}

\newcommand\varpm{\mathbin{\vcenter{\hbox{%
  \oalign{\hfil$\scriptstyle+$\hfil\cr
          \noalign{\kern-.3ex}
          $\scriptscriptstyle({-})$\cr}%
}}}}
\newcommand\varmp{\mathbin{\vcenter{\hbox{%
  \oalign{\hfil$\scriptstyle-$\hfil\cr
          \noalign{\kern-.3ex}
          $\scriptscriptstyle({+})$\cr}%
}}}}

\def\scititle{
Bridging DNP and MAS NMR dipolar recoupling - from static single crystal to spinning powders
}
\title{\bfseries \boldmath \scititle}

\author{
	José P. Carvalho$^{\dagger}$,
	Anders Bodholt Nielsen$^{\dagger}$,
	Enik\H{o} Baligács, Nino Wili, \\ and
    Niels Chr. Nielsen$^{\ast}$
 \and
	\small Interdisciplinary Nanoscience Center (iNANO) and Department of Chemistry, Aarhus University, 
    \\ \small Gustav Wieds Vej 14, DK-8000 Aarhus C, Denmark.
	\small$^\ast$Corresponding author. Email: ncn@chem.au.dk\and
	\small$^\dagger$These authors contributed equally to this work.
}

\begin{document} 

\maketitle

\begin{abstract} \bfseries \boldmath
Spin engineering of advanced pulse sequences has had a transformative impact on the development of nuclear magnetic resonance (NMR) spectroscopy, to an extending degree also electron paramagnetic resonance (EPR), and the hybrid between the two, dynamic nuclear polarization (DNP). Based on a simple formalism, we demonstrate that ($i$) single-crystal static-sample optimisations may tremendously ease design of experiments for rotating powders and ($ii$) pulse sequences may readily be exchanged between these distinct spectroscopies. Specifically, we design broadband heteronuclear solid-state NMR magic-angle-spinning (MAS) dipolar recoupling experiments based on the recently developed PLATO (PoLarizAtion Transfer via non-linear Optimization) microwave (MW) pulse sequence optimized on a single crystal for powder static-sample DNP. Using this concept, we demonstrate design of ultra-broadband $^{13}$C-$^{15}$N and $^2$H-$^{13}$C cross-polarization experiments, using PLATO on the $^{13}$C radio-frequency (RF) channel and square/ramped or RESPIRATION (Rotor Echo Short Pulse IRrAdiaTION) RF irradiation on the $^{15}$N and $^2$H RF channels, respectively. 
\end{abstract}

\newpage
\noindent
\section*{INTRODUCTION}
Over the past decades, coherent spectroscopy has undergone tremendous evolution by development of advanced instrumentation allowing for phase coherent pulse/waveform implementation supported by advanced pulse engineering methods. This applies not least in magnetic resonance, with nuclear magnetic resonance (NMR)\cite{Ernst_book} and magnetic resonance imaging (MRI) paving the way, later followed by pulsed electron paramagnetic resonance (EPR) \cite{Jeschke_book} and lately pulsed dynamic nuclear polarization (DNP) \cite{Slichter_DNP,Becerra_DNP,Corzelius_DNP_review}. Notable pulse engineering tools include average Hamiltonian theory (AHT) \cite{AHT,scBCH}, Floquet theory \cite{Weintraub,IVANOV202117}, Exact Effective Hamiltonian Theory (EEHT) \cite{EEHT}, and more recently Single-Spin-Vector Effective Hamiltonian Theory (SSV-EHT) \cite{shankar,SSV-EHT,SVEHT_EEHT}, all playing different roles in the development and understanding of increasingly advanced pulse sequences unraveling key information on the (spin) systems subject to investigation. Considering the aspect of systematic design of methods, in particular AHT and recently SSV-EHT single out by providing very direct insight into important determinants for systematic design, as will also be central in this article.  We will address the fundamentally important aspect that advanced experiments in NMR and EPR, static or rotating samples, single crystals and powders may easily be unified already at the stage of experiment design. While earlier addressed mainly in relation to specific solid-state NMR inspiration to pulsed DNP \cite{XiX_DNP,Uni_CP_DNP,TPPM_DNP}, this has a much wider consequence, as will be an objective of this paper. This applies to the ready exchange of methods between spectroscopies and for simplifying the task of method development which naturally would appear easiest for a single crystal (rather than powders representing many crystals which need to be coped with in parallel) and static samples (rather than rotating samples inducing another layer of time dependency). 

In this work, we will with specific focus on solid-state NMR and pulsed DNP theoretically and experimentally demonstrate two important, yet surprising, aspects that have not been addressed nor demonstrated during the many years of spin engineering in magnetic resonance. The first aspect is that development and optimization of pulse sequences for systems with anisotropic (i.e., orientation dependent) spin interactions for powder samples may not need to address spatial distribution over many independent orientations of molecules (crystallites) relative to the external magnetic field, and may not need to take into account sample spinning at first instance. The second aspect is that methods may easily be exchanged between static DNP and magic-angle-spinning solid-state NMR spectroscopy. The latter is not an entirely new line of thinking, but we demonstrate here an example that a highly advanced pulse sequence with great benefit may be translated from static DNP to MAS dipolar recoupling NMR with the remarkable aspect of reaching broadband performance in MAS NMR that has not yet proven possible to reach with current design strategies. The specific pulse sequence we use as origin to demonstrate these aspects is the so-called PLATO (PoLarizAtion Transfer via non-linear Optimization) pulse sequence recently proposed for broadband static-powder DNP \cite{PLATO_adv}. This single-spin microwave (MW) irradiation pulse sequence is here translated to two-spin radio-frequency (RF) irradiation mediated magic-angle-spinning (MAS) dipolar recoupling to form the $^{\text{PLATO}}$CP broadband variant to the double-cross-polarization \cite{DCP} experiment for $^{15}$N-$^{13}$C polarization transfer and the $^{\text{RESPIRATION-PLATO}}$CP experiment as a broadband variant to the RESPIRATION (Rotor Echo Short Pulse IRrAdiaTION mediated cross polarization) cross-polarization (CP) experiment \cite{RESPIRATIONCP,adRESPIRATION,2H_RESPIRATION} for $^2$H to $^{13}$C polarization transfer.

\section*{RESULTS}

While we could have started from static solid-state NMR - as the objective is to design MAS NMR experiments - we chose the more challenging start from a static DNP pulse sequence and recast this to two different solid-state NMR MAS CP experiments to demonstrate our two key findings. The theory is kept short in the main text, an extended account can be found in the Supplementary Material. 

Consider an electron (S) - nuclear (I) spin-pair subject to an external magnetic field and pulsed MW irradiation with the spin dynamics governed by the Hamiltonian 
\begin{equation}
  \mathcal{H}(t)=\omega_I  I_z+\Delta\omega_S  S_z+A S_z I_z+B S_z I_x+\mathcal{H}_\text{MW}(t) ,
  \label{Eq:1}
\end{equation}  
where $\Delta\omega_\text{S}=\omega_\text{S}-\omega_\text{MW}$ denotes the MW offset frequency, $\omega_S$ and $\omega_I$ the electron and nuclear Larmor frequencies; all in angular frequencies. $\mathcal{H}_\text{MW}(t)$ describes the MW pulse sequence with irradiation at the MW carrier frequency $\omega_{MW}$. $S_i$ and $I_i$ ($i$ $\in$ $x,y,z$) represent spin operators for the electron and nuclear spins.  $A$ and $B$  denote amplitudes for the secular and pseudo secular hyperfine coupling, respectively. Relative to DNP transfer of polarization from electron to nuclear spins, this Hamiltonian has  the intrinsic problem that the nuclear Larmor term (first term in Eq. (\ref{Eq:1})) modulates and effectively averages out the effect of the pseudosecular coupling term (fourth term in Eq. (\ref{Eq:1})) being the only source to DNP in cases without RF irradiation on the nuclear spins. In this case, the secular term is irrelevant for polarization transfer. 

\subsection*{Shaping the DNP Hamiltonian}

\begin{figure}[hbt!]
\centering
\includegraphics[]{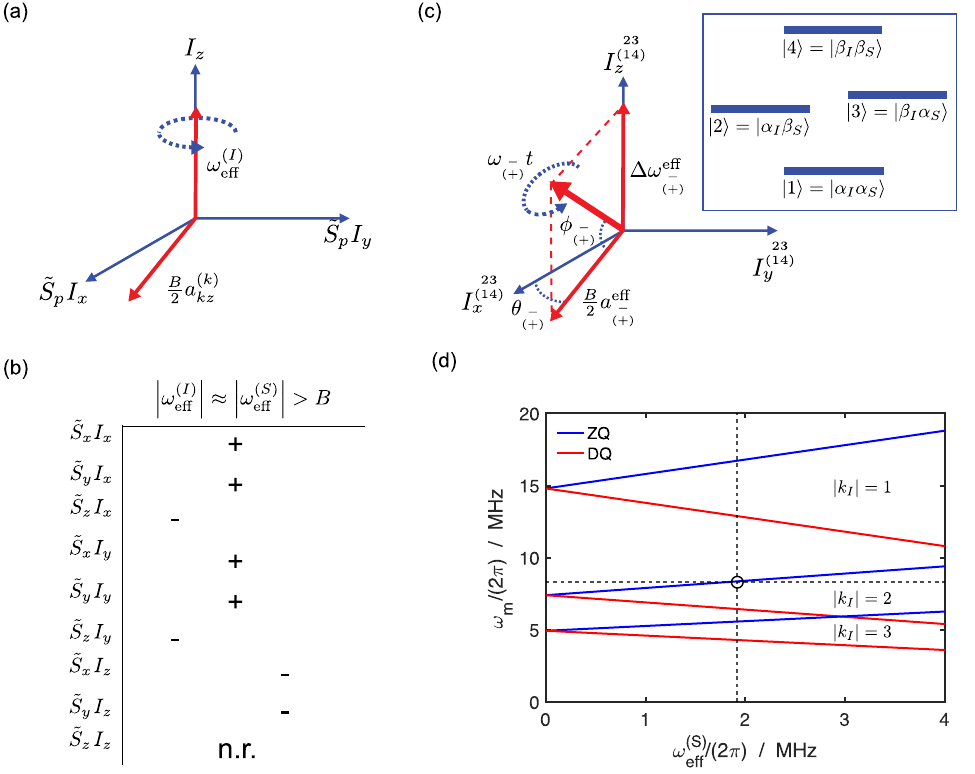}
\caption{\scriptsize  \textbf{Principle of averaging, resonances, and dipolar recoupling.} Illustration of averaging and recoupling of operator components for an electron-nuclear spin-pair  in DNP experiments. (a) Definition of the effective linear field and illustration of its impact on transverse elements of bilinear operators ($p$ = $x$ or $y$; $I$ and $\tilde{S}$ may be exchanged). (b) Listing of bilinear terms averaged (marked by "-"), potentially recoupled (marked by ``+''), and non-relevant (marked by ``n.r.'', applying to the secular hyperfine coupling in absence of RF irradiation on the nuclear spins). (c) Effective fields in ZQ (marked by 2-3 energy-level operators) and DQ (marked by 1-4 energy-level operators) invariant operator subspaces relevant for DNP (see insert). (d) ZQ (blue) and DQ (red) resonance conditions as function of the S-spin effective field, the MW pulse sequence modulation frequency, and the Fourier number $k_I$. The circle marks the ZQ resonance of the PLATO DNP experiment.
}
\label{fig:1}
\end{figure}

Averaging of the pseudo secular coupling may be prevented by modulating the MW irradiation with a frequency effectively matching the nuclear Larmor frequency - which may be referred to as dipolar recoupling. This modulation generates an effective field $\omega^{(S)}_\text{eff}$ along $\tilde{S}_z$ for the electron spin(s) in the interaction frame of the MW pulse sequence (marked by $\tilde{ }$ ) along with the already existing effective field $\omega^{(I)}_\text{eff}=\omega_I-k_I \omega_m$  along $I_z$ for the nuclear spin(s). $\omega_m$ is the modulation frequency of the pulse sequence (related to the duration $\tau_m$ of the MW pulse sequence element as $\omega_m = 2 \pi/\tau_m$), and $k_I$ an integer Fourier number (see further details in Supplementary Materials). These fields, illustrated in Fig. \ref{fig:1}a, modulate the bilinear (two-spin) terms in Eq. (\ref{Eq:1}). In the frame of the effective fields the only single-spin terms are $I_z$ and $\tilde{S}_z$, and provided reasonably sized, these effective fields averages all bilinear operators with only one transverse operator. This leaves 4 out of 9 bilinear operators potentially active for DNP as illustrated in Fig. \ref{fig:1}b with ``-'' denoting averaging, ``+" potential recoupling, and ``n.r.'' not relevant.

In the weak coupling regime (i.e., $|A|, |B| << |\omega_I|$), 
the effective Hamiltonian to first order governing DNP  may be derived using SSV-EHT \cite{SVEHT_EEHT,BEAM} (cf. Supplementary Materials) and formulated as 
\begin{equation}
\overline{\tilde{\mathcal{H}}}^{(1)}_{\substack{\scriptscriptstyle{ZQ}\\ \scriptscriptstyle{(DQ)}}}=\frac{B}{2} [a^{\varmp} I_x^{\substack{\scriptscriptstyle{23}\\ \scriptscriptstyle{(14)}}} +b^{\varmp}  I_y^{\substack{\scriptscriptstyle{23}\\ \scriptscriptstyle{(14)}}}] + \Delta\omega_{\varmp}^\text{eff}  I_z^{\substack{\scriptscriptstyle{23}\\ \scriptscriptstyle{(14)}}} ,
  \label{Eq:2}
\end{equation}
where we introduced the fictitious spin-1/2 operators \cite{fictitious_wokaun_ernst,fictitious_vega}  $I_x^{\substack{\scriptscriptstyle{23}\\ \scriptscriptstyle{(14)}}}=\tilde{S}_x I_x \varpm \tilde{S}_y I_y$, $I_y^{\substack{\scriptscriptstyle{23}\\ \scriptscriptstyle{(14)}}}=\tilde{S}_y I_x \varmp\tilde{S}_x I_y$, and $I_z^{\substack{\scriptscriptstyle{23}\\ \scriptscriptstyle{(14)}}}=\frac{1}{2}(\tilde{S}_z \varmp I_z)$, and defined  $\Delta\omega_{\varmp}^\text{eff}=-(\omega^{(I)}_\text{eff} \varpm \omega^{(S)}_\text{eff}$). We use here a notation with the upper sub/super-scripts referring to ZQ recoupling, while the lower ones in parentheses refer to DQ recoupling. The coefficients   $a^{\varmp}$ and $b^{\varmp}$ relate to pulse sequence specific interaction-frame Fourier coefficients as described in Supplementary Materials. The linear longitudinal fields allow for selection of either ZQ or DQ operation using pulse sequences with either $|\Delta\omega_-^\text{eff}| < |B|$ and $|\Delta\omega_+^\text{eff}| >> |B|$ or 
$|\Delta\omega_+^\text{eff}| < |B|$ and  $|\Delta\omega_-^\text{eff}| >> |B|$, respectively. The large linear field component  truncates the DQ Hamiltonian in the first case, and the ZQ Hamiltonian in the last case. The two phase components in Eq. (\ref{Eq:2}) may be described as an effective field $\frac{B}{2}$ scaled by $a_{\varmp}^\text{eff}=\sqrt{(a^{\varmp})^2+(b^{\varmp})^2}$ along, $I_x^{\substack{\scriptscriptstyle{23}\\ \scriptscriptstyle{(14)}}}$ rotated by an angle (azimuth) $\theta_{\varmp}=\arctan(b^{\varmp}/a^{\varmp})$ around $I_z^{\substack{\scriptscriptstyle{23}\\ \scriptscriptstyle{(14)}}}$. The remaining offset term tilts the effective field away from the transverse plane by an angle (zenith) $\phi_{\varmp}=\arctan(\frac{B}{2}\,a_{\varmp}^\text{eff}/\Delta \omega_{\varmp}^\text{eff})$ to give an effective field  $\omega_{\varmp}=\sqrt{(\frac{B}{2}\,a_{\varmp}^\text{eff})^2+(\Delta \omega_{\varmp}^\text{eff})^2}$. This leads to the Hamiltonian
\begin{equation}
  \overline{\tilde{\mathcal{H}}}_{\substack{\scriptscriptstyle{ZQ}\\ \scriptscriptstyle{(DQ)}}}= \omega_{\varmp}\left[
e^{-i \theta I_z^{\substack{\scriptscriptstyle{23}\\ \scriptscriptstyle{(14)}}}}\left( e^{-i \phi I_y^{\substack{\scriptscriptstyle{23}\\ \scriptscriptstyle{(14)}}}} I_x^{\substack{\scriptscriptstyle{23}\\ \scriptscriptstyle{(14)}}} e^{i \phi I_y^{\substack{\scriptscriptstyle{23}\\ \scriptscriptstyle{(14)}}}} \right)e^{-i \theta I_z^{\substack{\scriptscriptstyle{23}\\ \scriptscriptstyle{(14)}}}} \right],
  \label{Eq:3}
\end{equation}
with effective fields and angles illustrated in Fig. ~\ref{fig:1}c here formulated as two intertwined rotations. 

\subsection*{DNP Transfer Function}

DNP transfer from electron spin polarization $\tilde{S}_z=I_z^{14}+I_z^{23}$ to nuclear spin polarization $I_z=I_z^{14}-I_z^{23}$ may be obtained by a ZQ operation inverting $I_z^{23}$ (with $I_z^{14}$  invariant) leading to net positive transfer, or correspondingly by a
DQ operation  inverting $I_z^{14}$ (with $I_z^{23}$ invariant) leading to net negative transfer.  Using Eq. (\ref{Eq:3}), the  efficiency of the polarization transfer may readily be evaluated as 
\begin{eqnarray}
  \langle  I_z\rangle_{\substack{\scriptscriptstyle{ZQ}\\ \scriptscriptstyle{(DQ)}}} (t) &=& 
  \varpm\langle  \rho_0| \tilde S_z\rangle \cos^2(\phi_{\varmp})\sin^2\left(\frac{1}{2}\omega_{\varmp} t\right)
    \nonumber \\
 &=&
 \varpm \langle  S_z| \tilde S_z\rangle \frac{(B^\text{eff})^2}{(B^\text{eff})^2+(\Delta \omega_{\varmp}^\text{eff})^2} \sin^2\left[\frac{t}{2} \sqrt{(B^\text{eff})^2+(\Delta \omega_{\varmp}^\text{eff})^2}\right],
  \label{Eq:4}
\end{eqnarray}
introducing the effective scaled hyperfine coupling constant $B^\text{eff}=\frac{B}{2}\, a_{\varmp}^\text{eff}$ and  $\rho_0 = S_z$ representing the initial density operator. This expression reduces to 
  $\langle  I_z\rangle_{\substack{\scriptscriptstyle{ZQ}\\ \scriptscriptstyle{(DQ)}}} (t) =
 \varpm \langle  \rho_0| \tilde S_z\rangle \sin^2(\frac{t}{2}B^\text{eff})$ 
upon matching resonance. Equation (\ref{Eq:4}) reveals the ingredients needed for efficient DNP experiments being broad-banded with respect to the electron spin offset: The pulse sequence needs to provide ($i$) a constant linear field $\Delta \omega_{\varmp}^\text{eff}$ over the targeted offset profile, ($ii$) a high scaling factor $a_{\varmp}^\text{eff}$ to provide fast transfer (thereby reducing competition from other interactions and relaxation), and ($iii$) good separation of the ZQ and DQ Hamiltonians both of which are present at the same time  with proportions  depending on the modulation time $\tau_m=2\pi/\omega_m$ and the scaling factors $a_{\varmp}^\text{eff}$. 

From Eq. (\ref{Eq:4}), it is evident that it is important to select or optimize pulse sequences which appropriately separate ZQ and DQ transfers as these two, through different sign of transfer, may interfere destructively, e.g., for powder samples or under conditions of inhomogeneous MW fields. The ZQ and DQ transfer maximizes at $\omega_I-k_I \omega_m + \omega_\text{eff}^{(S)}$ = 0 and $\omega_I-k_I \omega_m - \omega_\text{eff}^{(S)}$ = 0, respectively,  exposing the important role of the modulation frequency of the pulse sequence. Plotting the ZQ (blue) and DQ (red) resonance conditions, Fig. \ref{fig:1}d reveals that a good separation may be obtained for a sequence of length 120 ns (corresponding to a modulation frequency of 8.33 MHz; values inspired from the PLATO sequence \cite{PLATO_adv} originally developed for pulsed DNP at X-band frequency) requiring an electron spin effective field in the order of $\pm$1.9 MHz (marked by dashed lines) being compatible with the selection rules above. The crossing of the dashed lines, marked by an open circle, in Fig. \ref{fig:1}d represents good choices for a ZQ recoupling sequences with minimal interference from DQ recoupling. We note that the alternative more well-separate matching conditions at higher modulation frequencies may be less attractive as the corresponding shortening of the pulse sequence element renders flexibility to stabilize the linear field more challenging. It is also important to consider the effective fields relative to the size of the hyperfine coupling to ensure proper averaging of unwanted bilinear components (vide supra). 

\subsection*{Single crystal vs powder optimization}

An important aspect to consider is the dependency of the transfer (or FOM, Figure Of Merit) function in Eq. (\ref{Eq:4}) on the relative orientation of the electron-nuclear spin axis and the external field axis as characterized by an angle $\beta_{PL}$. This dependency enters via the pseudo-secular hyperfine coupling constant, $B = \frac{3}{2}T \sin(2 \beta_{PL})$, with corresponding scaling of $B^\text{eff}$ in Eq. (\ref{Eq:4}). $T = -\gamma_I \gamma_S \hbar \frac{\mu_0}{4 \pi} \frac{1}{r_{IS}^3}$ is the anisotropy of the hyperfine coupling in the point dipole model. We note that the corresponding scaled secular hyperfine coupling constant, $A = T (3 \cos^2(\beta_{PL})-1)$ does not enter as source of recoupling provided irradiation only occurs on the MW channel. With reference to Fig. \ref{fig:1}c, orientational scaling of the pseudo-secular coupling corresponds to variations in the pulse sequence induced nutation around the ZQ/DQ effective field axis by angle $\omega_{\varpm} t$. 

\begin{figure}[ht]
\centering
\includegraphics[scale=1.0]{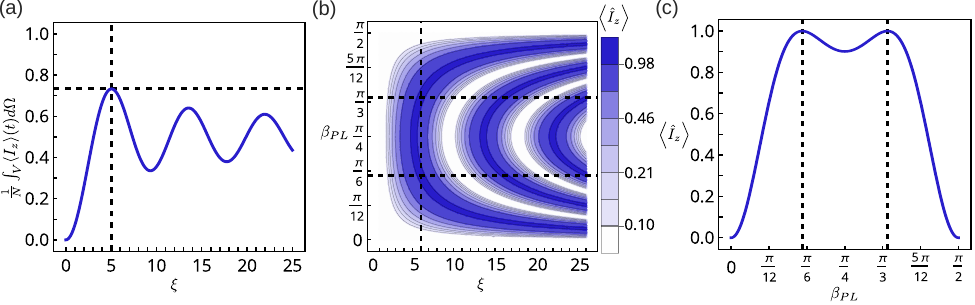}
\caption{\scriptsize  \textbf{Single crystal versus powder averaging.} Numerical analysis of the impact of orientational averaging on the DNP transfer efficiency under conditions of a matched ZQ resonance. (a) Orientationally averaged DNP transfer efficiency as a function of $\xi=T a_{\protect\varmp}^\text{eff} t $, calculated using Eq. (\ref{Eq:6}), truncated at $j=50$. The dotted lines highlight the global maximum of the function for which the impact of orientational averaging is minimized. (b) Contour plot showing the dependency of the FOM (\textit{cf.} Eq. (\ref{Eq:4}) for a ZQ resonance condition, $\Delta \omega_{\protect\varmp}^\text{eff})=0$ and $\langle  \rho_0| \tilde S_z\rangle =1$) on $\beta_{PL}$ and $\xi$. Due to the symmetry of $\sin(2\beta_{PL})$, we only consider the $0$ to $\pi/2$ range of $\beta_{PL}$ angles. The vertical dotted line mark $\xi\approx5.04$ corresponding to the global maximum shown in (a), while the horizontal dotted lines mark the values of $\beta_{PL}$ for which the polarization transfer is complete, i.e., $\langle  I_z\rangle=1$. (c) Vertical slice of the contour plot in (b) for $\xi\approx5.04$. The vertical dotted lines mark the values of $\beta_{PL}$ for which the polarization transfer is complete.
}
\label{fig:2}
\end{figure}

\sloppy
For a powder sample,  Eq. (\ref{Eq:4})  needs to be evaluated with orientational averaging $\frac1N\int_V\langle I_z\rangle_{\substack{\scriptscriptstyle{ZQ}\\\scriptscriptstyle{(DQ)}}} (t) d\Omega$, where $N$ is the normalization constant and $V$ is the integration volume defined by the range of involved Euler angles. Upon matching  a pure ZQ (or DQ) resonance condition, the powder-averaged DNP transfer efficiency may be expressed by the integral
\begin{equation}
\frac1N\int_V\langle I_z\rangle_{\substack{\scriptscriptstyle{ZQ}\\\scriptscriptstyle{(DQ)}}} (t) d\Omega  =
 \varpm \frac14\langle  \rho_0| \tilde S_z\rangle \int_0^\pi\left\{1-\cos\left[\frac34 \xi \sin(2 \beta_{PL})  \right]\right\}\sin(\beta_{PL})d\beta_{PL},  \label{Eq:5}
\end{equation} 
where $\xi=T a_{\varmp}^\text{eff} t$.  Employing the Jacobi-Anger expansion \cite{abramowitz1968handbook,mueller1995analytic,herzfeld1980sideband}, this can be recast as \begin{equation}
\frac1N\int_V\langle I_z\rangle_{\substack{\scriptscriptstyle{ZQ}\\\scriptscriptstyle{(DQ)}}} (t) d\Omega =
 \varpm \frac12\langle  \rho_0| \tilde S_z\rangle\left[1+J_0\left(\frac34 \xi \right)+\sum_{j=1}^\infty \frac{2 }{16j^2-1} J_{2j}\left(\frac34 \xi \right)\right], \label{Eq:6}
\end{equation}
where $J_j$ correspond to  Bessel functions of the first kind (verified with numerical simulations in Fig. \ref{fig:S1}, Supplementary Materials, also containing a more detailed derivation of the equations above). In this formulation, it is clear that upon matching a DNP resonance condition, the effect of the orientational averaging is determined solely by the variable $\xi$ being the product of the mixing time, the hyperfine coupling anisotropy, and the pulse sequence specific scaling factor. Figure \ref{fig:2}a shows the dependency of the powder-averaged DNP transfer efficiency on $\xi$.  The impact of orientational averaging is minimized  at $\xi\approx5.04$ where $\frac1N\int_V\langle I_z\rangle_{\substack{\scriptscriptstyle{ZQ}\\\scriptscriptstyle{(DQ)}}} (t) d\Omega \approx0.73$, corresponding to the first maximum of the curve, as marked with dashed lines. For further insight, Fig. \ref{fig:2}b illustrates the dependency of the single-crystal FOM for a matched resonance, corresponding to the integrand in Eq. (\ref{Eq:5}), on both $\xi$ and $\beta_{PL}$. For $\xi\approx 5.04$, it can be seen that there are two orientations at $\beta_{PL}\approx0.49$ and $\beta_{PL}\approx1.08$ (corresponding to 28.07$^\circ$ and 61.88$^\circ$, respectively) for which the  polarization transfer reaches unity. The two maxima are clearly visible in Fig. \ref{fig:2}c, representing a slice from the contour at $\xi\approx5.04$. At glance, this suggests that one of these two specific orientations should be chosen for "powder-independent" pulse sequence optimization, since it can then be ensured that the impact of the orientational averaging is minimized. However, for a given scaling factor (defined by the pulse sequence) and anisotropy of the hyperfine interaction (defined by the spin system), the mixing time can, within the discretization of the modulation time, \textit{a posteriori} be adjusted so that the decrease in overall transfer is minimized. In that sense, the choice of orientation is not significant as long as we are able to establish a single ZQ/DQ resonance with complete polarization transfer, \textit{i.e.} $\langle  I_z\rangle=1$. However, this does not hold if one selects an orientation zeroing the hyperfine coupling, thereby preventing the establishment of a resonance condition, or an orientation leading to a small pseudo-secular coupling, as it may fail to ensure convergence of the first-order effective Hamiltonian for all orientations. 

This result can be extended to a range of offsets provided the pulse sequence is optimized to yield a constant linear field (i.e., a very small mismatch of the effective fields over the range of offsets) and a stable scaling factor, that the initial density operator is aligned with the effective field. If these conditions are met, powder averaging will only result in a uniform decrease in the overall transfer efficiency for that range of offsets. From a pulse sequence engineering point of view, this implies that to optimize pulsed DNP sequences, only a single crystal orientation is needed to steer the optimization.  This applies as long as the first-order  effective Hamiltonian  description is valid (\textit{vide supra}). Importantly, an equivalent formulation can be found for the secular coupling, as elaborated on in Supplementary Material. This implies that static-powder (and thereby also, as will be shown below, MAS) solid-state NMR experiments can be optimized based on a static-sample single-crystal orientation, in this case with $\beta_{PL}$ differing from the magic angle. This is a notable result as it significantly reduces the complexity of optimization, the computational speed, and increase the base for understanding the spin dynamics underlying resulting recoupling methods.

\subsection*{From static DNP to MAS dipolar recoupling}

Above, we have focused on DNP experiments conducted using MW irradiation at the electron spin(s) to promote dipolar recoupling through matching to the nuclear Larmor frequency. This situation may readily be reformulated to MAS solid-state NMR dipolar recoupling in an I - S nuclear spin-pair system with the aim of driving polarization transfer from I to S. Using an DNP-inspired pulse sequence on the S spin, we match the effective field on the S spin ($\omega_\text{eff}^{(S)}$) to an RF field with constant amplitude $\omega_I^\text{RF}$ on the I spin and the spinning frequency $\omega_r$ as
\begin{equation}
\omega_I^\text{RF} - k_I \omega_m \pm \omega_\text{eff}^{(S)} = n \omega_r , \label{Eq:7}
\end{equation}
 where $\omega_m$ is the modulation frequency of the pulse sequence, $k_I$ and $n$ are integers, and with $\pm$ invoking ${\substack{\scriptscriptstyle{ZQ}\\ \scriptscriptstyle{(DQ)}}}$ dipolar recoupling. This extremely simple translation obviously also applies in the case of translation to static-sample solid-state NMR by setting $\omega_r=0$. We note this resonance condition bears close resemblance to earlier Hartmann-Hahn like resonance conditions with the subtle detail that it includes the effective field of the pulse sequence on the S spins, which in the simple case of continuous wave CP is just the amplitude of a constant field. Here it is generalized to any pulse sequence for which readily can calculate the effective field. We should also note that the same trick may obviously be used on the I spins through replacement of $\omega_I^\text{RF}$ with $\omega^{(I)}_\text{eff}$ for a more complicated pulse sequence.

\subsection*{$^{15}$N-$^{13}$C $^{\text{PLATO}}$CP cross polarization}

To demonstrate the transfer of single-crystal optimized broadband static DNP pulse sequence to a broadband cross polarization experiment for $^{15}$N (I) to $^{13}$C (S) transfer under fast MAS conditions based on the matching condition in Eq. (\ref{Eq:5}), we take origin in the recently developed PLATO static-powder pulsed DNP  sequence \cite{PLATO_adv}. This sequence, optimized for a single crystallite with angle $\beta_{PL} \approx 64.9^\circ$, consists of 24 pulses of length 5 ns, different phases, and peak amplitude 32 MHz (details in Materials and Methods). To arrive at NMR  conditions, we expand the time axis of the experiment by a factor of 1000 (going from ns to $\mu$s) and reduce the pulse amplitude  by a factor of 1000 (going from MHz to kHz). In this setting, we obtain ZQ dipolar recoupling under the condition of $\omega_r/(2\pi)$=25 kHz spinning using an RF field with amplitude $|\omega_I^\text{RF}/(2\pi)|$ = 10.224 kHz on the $^{15}$N (I) spins selecting the $k_I=-2$ and $n=1$ resonance (among several solutions, see Fig. \ref{fig:1}d and Eq. (\ref{Eq:5})) and the PLATO-specific linear field of $\omega_\text{eff}^{S}/(2 \pi)$ = -1.911 kHz. Through scaling of all S-spin parameters by 1000, we anticipate that this specific recoupling experiment will provide an off-resonance bandwidth of around 80 kHz for  $^{13}$C  as predicted from the corresponding 80 MHz for  PLATO DNP  on  electron spins \cite{PLATO_adv}. With this experiment representing a broadband alternative to the widely used cross-polarization (CP) experiment \cite{CP}, or in the present context of transfer between $^{15}$N and $^{13}$C the so-called Double Cross Polarization (DCP) experiment \cite{DCP}. Due to its generality, we henceforth refer to this experiment as $^{\text{PLATO}}$CP.

\begin{figure}[H]
\centering
\includegraphics[scale=1.0]{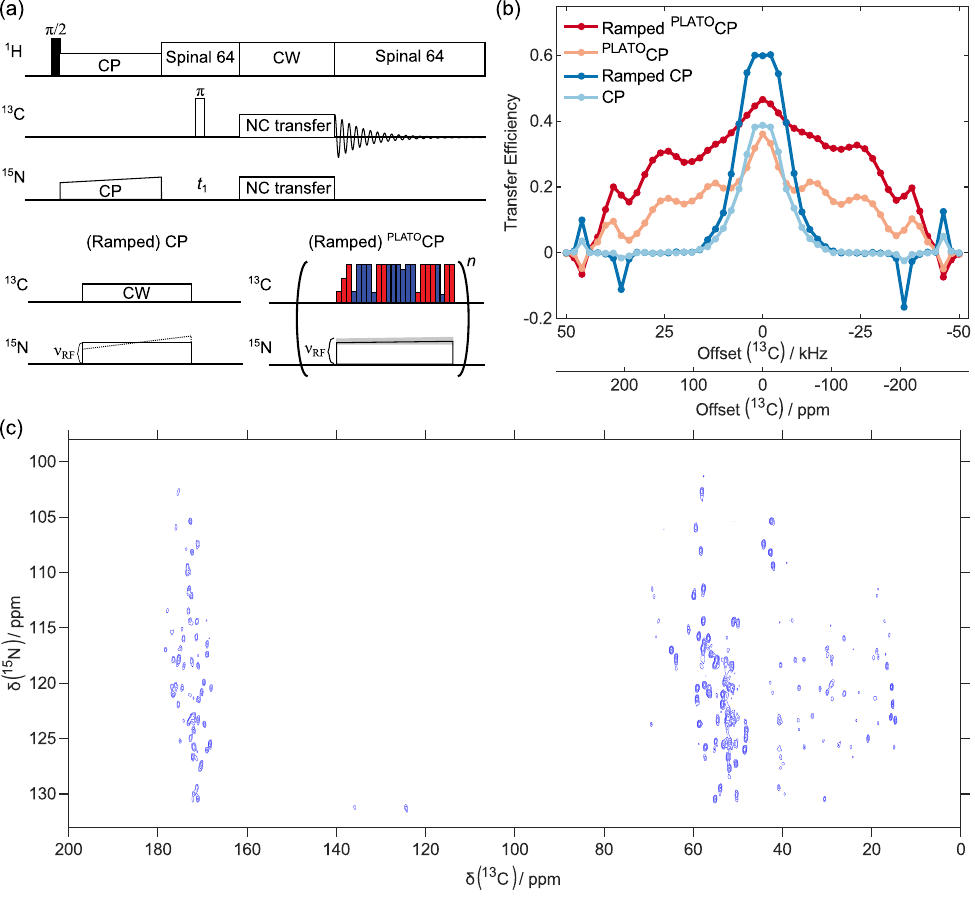}
\caption{\scriptsize  \textbf{$^{\text{PLATO}}$CP and DCP pulse sequences and experimental demonstrations.} Experimental evaluation of DCP (panel a, lower left) and $^{\text{PLATO}}$CP (panel a, lower right) dipolar recoupling experiments and ramped variants incorporated into the pulse sequence in (panel a, top) for $^{15}$N $\rightarrow$ $^{13}$C$_\alpha$ transfer in rotating powder samples.  (b) Comparison of the S-spin ($^{13}$C) resonance offset profile of DCP (1.6 ms mixing time,  $^{13}$C and $^{15}$N RF field strengths of 32.0 and 4.4. kHz, respectively), ramped-DCP (7.0 ms mixing time,  $^{13}$C and $^{15}$N RF field strengths of 32.0 and 4.8 (70-100\% RAMP) kHz, respectively), $^{\text{PLATO}}$CP (2.4 ms mixing time,  $^{13}$C and $^{15}$N RF field strengths of 32.0 kHz (max. amplitude; 120 $\mu$s PLATO sequence) and 9.9 kHz, respectively), and a ramped $^{\text{PLATO}}$CP (6.6 ms mixing time,  $^{13}$C and $^{15}$N RF field strengths of 32.0 kHz (max. amplitude; 120 $\mu$s  PLATO sequence) and 10.6 (90-100\% RAMP) kHz, respectively) for a $^{13}$C$_\alpha$,$^{15}$N-labeled  sample of glycine. (c) 2D $^{13}$C-$^{15}$N correlation spectrum of uniformly  $^{13}$C,$^{15}$N-labeled GB1 obtained using ramped $^{\text{PLATO}}$CP (parameters as above) for $^{15}$N $\rightarrow$ $^{13}$C polarization transfer. The spectra were recorded at 16.4 T (700 MHz for protons) using 25 kHz sample spinning. Further experimental details are given in Materials and Methods.}
\label{fig:3}
\end{figure}

Using powder samples of $^{13}$C$_\alpha$,$^{15}$N and $^{15}$N,$^{13}$C labeled  glycine and the 56-residue protein GB1 (B1 immunoglobulin binding domain of streptococcal protein G), the performance of $^{\text{PLATO}}$CP and ramped $^{\text{PLATO}}$CP are evaluated relative to  corresponding DCP and ramped DCP methods. The experiments were performed using a static magnetic field of 16.4 T (700 MHz for $^1$H) and 25 kHz sample spinning. The relevant pulse sequences are shown in Fig. \ref{fig:3}a, while Fig. \ref{fig:3}b compares the S ($^{13}$C) spin offset profiles of  $^{15}$N $\rightarrow$ $^{13}$C transfer for the four recoupling experiments. The plots in Fig. \ref{fig:3}b  clearly demonstrate the expected broadband profile of the $^{\text{PLATO}}$CP experiment approaching 80 kHz (or more than 400 ppm for $^{13}$C at 16.4 T), which to the best of our knowledge far exceeds the capabilities of previous recoupling experiments under similar conditions. The PLATO experiments were performed with a peak RF amplitude on $^{13}$C being 32 kHz (bandwidth $>$2.3 larger than the RF amplitude) matched to an RF field on the $^{15}$N RF channel of around 10 kHz experimentally. These parameters enable ready combination with efficient $^1$H decoupling typically needing to be in excess of 2-3 times the RF field amplitude on the low-$\gamma$ RF channels. Apart from demonstration of the substantial improvement in broadbandedness by using the PLATO variants relative to typically used DPC and ramped DCP, Fig. \ref{fig:3} also provide a direct comparison of transfer efficiencies. A marked improvement going from DCP to ramped-DCP is expected due to better compensation of RF inhomogeneity in the latter experiment. This also applies (although to lesser extent) to the $^{\text{PLATO}}$CP variants which in the present implementation only is compensated for RF inhomogeneity effects on the $^{13}$C RF channel as inherited from the optimized compensation of the DNP experiment \cite{PLATO_adv}.  Numerical simulations obtained using the open-source SIMPSON \cite{SIMPSON,SIMPSONint} software are given Fig. S2 (Supplementary Material). These, performed with parameters matching the experimental setup and a 5\% Lorentzian RF inhomogeneity, qualitatively support the experimentally observations. We note that the extreme broadbandedness of $^{\text{PLATO}}$CP is particularly interesting noting the steady increase in the magnetic fields of state-of-the-art NMR instrumentation  calling for design of pulse sequences performing at large bandwidths.

To demonstrate the dual-side $^{15}$N to $^{13}$C$_\alpha$ and  $^{13}$C' transfer capability of the broadband excitation profile, Figure \ref{fig:3}c shows a 2D $^{15}$N,$^{13}$C correlation spectrum obtained for an uniformly  $^{13}$C,$^{15}$N-enriched sample of GB1 obtained using 25 kHz sample spinning and ramped $^{\text{PLATO}}$CP for  $^{15}$N $\rightarrow$ $^{13}$C transfer with a mixing time of 7.8 ms. The spectrum clearly demonstrates the expected cross-peaks in both the aliphatic and the carbonyl region of the $^{13}$C dimension, which due to the broadbandedness of the $^{\text{PLATO}}$CP scheme easily may achieved even at the highest known NMR field conditions both using transient and adiabatic/ramped implementations.
\subsection*{$^{2}$H-$^{13}$C $^{\text{RESPIRATION-PLATO}}$CP cross polarization}

\begin{figure}[H]
\centering
\includegraphics[scale=1.0]{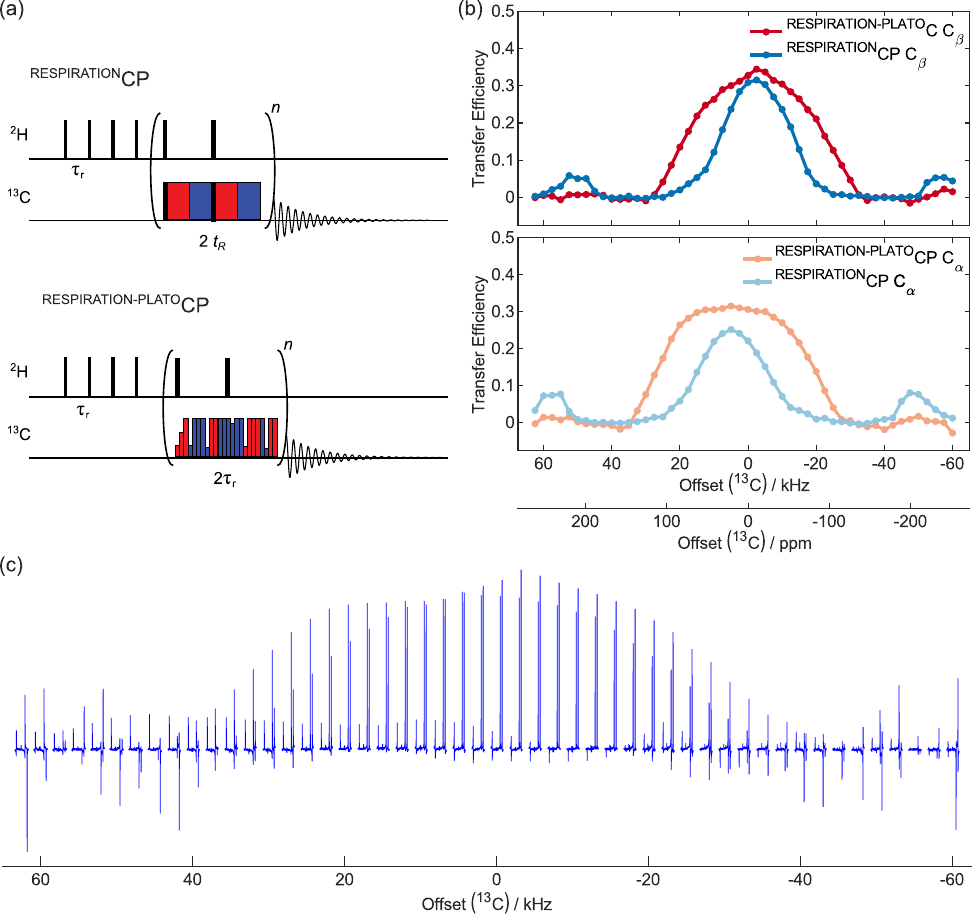}
\caption{\scriptsize \textbf{$^{\text{RESPIRATION}}$CP (upper panel) and $^{\text{RESPIRATION-PLATO}}$CP (lower panel) pulse sequences and experimental demonstrations.} (a) $^{\text{RESPIRATION}}$CP and $^{\text{RESPIRATION-PLATO}}$CP  pulse sequences and (b) experimental demonstration of $^{13}$C offset profile for $^{2}$H $\rightarrow$ $^{13}$C polarization transfer for a uniformly $^2$H,$^{13}$C,$^{15}$N-labeled sample of L-alanine recorded at 22.3 T (950 MHz for $^1$H) using 16.67 kHz spinning. The initial $(\pi/2)_x$ excitation on the $^2$H channel was accomplished using at RESPIRATION-4 pulse with each pulse separated by one rotor period ($\tau_r$ = 60 $\mu$s) and having a flip-angle of $\pi/8$ (1.44 $\mu$s; $\omega_{RF}/(2\pi$) = 43.5 kHz). The RESPIRATION sequence for polarization transfer used 4 $\mu$s $x$-phase $^2$H RF pulses with 39.4 kHz amplitude, while each of the two rotor periods of the $^{\text{RESPIRATION}}$CP element on the $^{13}$C RF channel used 4 $\mu$s RESPIRATION pulse and 28 $\mu$s pulses of phase $x$ and -$x$, all with an amplitude of 32.0 kHz. The $^{\text{RESPIRATION-PLATO}}$CP element used the same RF irradiation scheme at the $^2$H RF channel with 33.9 kHz, and on the $^{13}$C channel a 2$\tau_r$ = 120 $\mu$s PLATO sequence with maximum amplitude of 32.0 kHz. In both cases, the CP period was $10\tau_r$ = 600 $\mu$s. (c) Array of $^{\text{RESPIRATION-PLATO}}$CP spectra recorded with offset spanning from -60 kHz to +62.5 kHz and using a $^{\text{RESPIRATION-PLATO}}$CP period of $100\tau_r$ = 6.0 ms to allow observation also of the long-range $^2$H coupled carbonyl, otherwise parameters as in (b). }
\label{fig:4}
\end{figure}

To further illustrate the easy adaptivity and the versatility of single-crystal DNP based solid-state NMR MAS recoupling and the associated prospects of making ultra-broadband recoupling with low-modest RF power consumption, Fig. (\ref{fig:4}) demonstrates the PLATO DNP pulse sequence translated to $^{2}$H $\rightarrow$ $^{13}$C cross polarization for a powder sample of uniformly $^2$H,$^{13}$C,$^{15}$N- labeled L-alanine using 16.67 kHz MAS at 22.3 T (corresponding to 950 MHz for $^1$H). While attractive as a source to resolution enhancement in $^1$H-detected solid-state NMR experiments (through dilution of $^1$H spins), selective excitation of hydrophobic/hydrophilic domains of partially deuterated proteins, information about protein dynamics \cite{2H_dynamics_RESPIRATION_rienstra,2H_dynamics_akbey}, or as additional source of polarization for partially deuterated proteins \cite{QuadResMAS,RAPID}, it is well known that triple- and quadrupole resonance $^1$H-$^2$H-$^{13}$C(-$^{15}$N) experiments involving deuterons may be challenged by low  RF power in typical multiple-channel MAS probes relative to the required excitation bandwidth. This challenge may on the $^2$H channel be solved using so-called Rotor-Echo-Short-Pulse-IrRAdiaTION pulses (RESPIRATION) and cross polarization ($^{\text{RESPIRATION}}$CP) \cite{RESPIRATIONCP,adRESPIRATION} enabling good excitation using RF field strength as low as 10-30 kHz on the $^2$H RF channel, while providing some broadbandedness on the $^{13}$C channel \cite{2H_RESPIRATION}. 
Here we follow up on this challenge by replacing the RESPIRATION pulses and the associated $+x$ and $-x$ phase spin-lock pulses on the $^{13}$C RF channel with a DNP-inspired PLATO element (S spin) following the pulse sequence schematics in Fig. \ref{fig:4}a.

The conventional $^{\text{RESPIRATION}}$CP and the  proposed $^{\text{RESPIRATION-PLATO}}$CP pulse sequences shown in Fig. \ref{fig:4}a utilize first a RESPIRATION-4 $\pi/2$ excitation pulse \cite{OC_2H} followed by a  $^{2}$H $\rightarrow$ $^{13}$C CP element implemented either as $^{\text{RESPIRATION}}$CP \cite{RESPIRATIONCP,adRESPIRATION} or as $^{\text{RESPIRATION-PLATO}}$CP for direct comparison. In both cases, the linear field on $^2$H (I) spin is generated by a train of RESPIRATION pulses of length 4 $\mu$s and RF amplitudes of 39.4 and 33.9 for kHz for $^{\text{RESPIRATION}}$CP or $^{\text{RESPIRATION-PLATO}}$CP, respectively. This results in net effective fields of 2.63 and 2.26 kHz, both numerically close to the effective field of the PLATO sequence (-1.911 kHz). A peak RF field strength of 32.0 kHz is used for the $^{13}$C $^{\text{RESPIRATION}}$CP RF irradiation. In the  case of  $^{\text{RESPIRATION-PLATO}}$CP the $^{13}$C RF irradiation is based on the PLATO DNP pulse \cite{PLATO_adv} sequence with the same amplitude as above, but in this case, matched to a spinning frequency of 16.667 kHz.  From the plots in Fig. \ref{fig:4}b, it becomes apparent that the $^{\text{RESPIRATION-PLATO}}$C experiment is (slightly) more efficient and markedly more broadbanded on the $^{13}$C channel than the corresponding $^{\text{RESPIRATION}}$CP experiment for both $^{13}$C$_\alpha$ and $^{13}$C$_\beta$ with the directly attached deuterons characterized by quadrupolar coupling constants ($C_Q$) of 60 kHz and 153 kHz, respectively. The experimental findings are supported by numerical simulations in Fig. S3 (Supplementary Materials). Figure \ref{fig:4}c shows an array of $^{13}$C $^{\text{RESPIRATION-PLATO}}$CP spectra recorded with a longer mixing period (6.6 ms rather than 0.6 ms) to demonstrate excitation of both the directly bonded $^{13}$C$_\alpha$ and $^{13}$C$_\beta$ spins as well as the remotely bonded carbonyl $^{13}$C´ spin recorded at different carrier frequency offsets.

\section*{DISCUSSION}
In this work, we have demonstrated two essential new features of pulse sequence engineering in coherent spectroscopy. While much more general, we here have focused on the interplay between DNP in static powder samples and magic-angle-spinning solid-state NMR. We have demonstrated that dipolar recoupling pulse sequences optimized for single-crystal static samples (here DNP) cover optimal sequences for powder samples, and we have demonstrated that such sequences can be translated directly in to operation on powder samples under magic angle spinning conditions. A simple resonance equation provides easy translation between the different modalities of magnetic resonance. This uniqueness among related disciplines may provide important new understanding and insight in the basic principles of experimental methods. The possibility of developing and optimizing pulse sequences for single crystals in static samples tremendously simplifies the task of developing optimal experimental methods, but also has the stunning effect of providing sequences for MAS dipolar recoupling with broadband behavior not yet found by direct optimization under the conditions of rotating powders with averaging over three crystallite angles and handling sample rotation as opposed to single-crystallite static-sample optimization. 

We envisage that the specific experimental methods developed in this paper will find immediate useful applications in solid-state NMR of rotating samples through their unique broadbandedness, and that the design principles may offer a useful concept for more general and highly efficient design of methods for magnetic resonance and quantum sensing applications.

\section*{MATERIALS AND METHODS}

 The PLATO sequence, here in solid-state NMR implementation (amplitudes reduced by a factor of 1000 and timings expanded by a factor of 1000 relative to the DNP version \cite{PLATO_adv}) consists of 24 5-$\mu$s pulses of amplitude \{9.31, 20.36, 32.00, -9.29, -32.00, -32.00, -32.00, -7.86, 32.00, 32.00, -32.00, -32.00, -32.00, -28.04, -32.00, -32.00, 8.60, 32.00, 32.00, 32.00, -32.00, -6.94, 32.00, 32.00\} kHz. 

 The $^{15}$N-$^{13}$C $^{\text{PLATO}}$CP, ramped $^{\text{PLATO}}$CP, and corresponding DCP experiments were performed on a 16.4 T (700 MHz for $^1$H) Bruker Avance III HD wide-bore NMR spectrometer using a 1.3 mm HDCN quadruple-resonance MAS probe.  Experiments were performed for $^{15}$N-$^{13}$C$_\alpha$ and  uniformly-$^{15}$N-$^{13}$C labeled samples of glycine and the B1 immunoglobulin binding domain (GB1) of streptococcal protein G, respectively, using 25 kHz spinning and the pulse sequences shown in Fig. \ref{fig:3}a. The glycine offset spectra, forming basis for the integrated intensities shown in Fig. \ref{fig:3}b, were recorded using the pulse sequences in Fig. \ref{fig:3}a by accumulation of 8 transients with a relaxation delay of 3 s, omitting the t$_1$ evolution time, and a $^1$H $\pi/2$ pulse operating with an RF field strength of 71.4 kHz. The $^1$H $\rightarrow$ $^{15}$N CP used a mixing time of 2.3 ms and $^{1}$H and $^{15}$N RF field strengths of 79.9 and 37.0 (70-100\% RAMP) kHz, respectively. The $^{15}$N $\rightarrow$ $^{13}$C transfer with DCP was carried out using 1.6 ms of mixing time,  $^{13}$C and $^{15}$N RF field strengths of 32.0 and 4.4. kHz, respectively), while ramped-DCP employed 7.0 ms of mixing time, and $^{13}$C and $^{15}$N RF field strengths of 32.0 and 4.8 (70-100\% RAMP) kHz, respectively. $^{\text{PLATO}}$CP used 2.4 ms mixing,  $^{13}$C and $^{15}$N RF field strengths of 32.0 kHz (peak amplitude; 120 $\mu$s PLATO element) and 9.9 kHz, respectively. Ramped $^{\text{PLATO}}$CP used 6.6 ms of mixing time,  $^{13}$C and $^{15}$N RF field strengths of 32.0 kHz (peak amplitude; 120 $\mu$s  PLATO element) and 10.6 (90-100\% RAMP) kHz, respectively. SPINAL-64 \cite{spinal64} and continuous wave (CW) decoupling used RF field strengths of 100.0 and 125.0 kHz, respectively. The transfer efficiencies were calculated by integration of the spectral intensities, considering reference single $\pi/2$-pulse $^{13}$C and $^{15}$N spectra, with RF field strengths of 36.8 and 41.0 kHz and relaxation delays of 480 and 120 s, respectively, combined with $^1$H $\rightarrow$ $^{15}$N CP employing the same RF field strengths as the DCP experiments and a relaxation delay of 30 s. The 2D $^{15}$N,$^{13}$C correlation GB1 spectrum was acquired with a repetition delay of 3 s, 256 increments in the indirect dimension, spectral window of 50 ppm, and  160 scans per increment. An RF field strength of 64.4 kHz was used for the $^1$H $\pi/2$ pulse while  RF field strengths of 100.0 kHz (SPINAL-64) and 125.0 kHz (CW) were used for decoupling. For the $^1$H $\rightarrow$ $^{15}$N CP transfer, a mixing time of 1.0 ms was used with 84.9 kHz RF field strength for $^1$H and 38.1 kHz (70-100\% RAMP) for $^{15}$N. For the $^{15}$N $\rightarrow$ $^{13}$C ramped $^{\text{PLATO}}$CP transfer, a mixing time of 7.8 ms was used with  32.0 kHz RF field strength for $^{13}$C (peak amplitude; 120 $\mu$s PLATO element) and 10.3 kHz (90-100\% RAMP) for $^{15}$N. A  $\pi/2_x-\pi_y-\pi/2_x$ composite pulse $\pi$ was used for $^{13}$C refocusing during the $t_1$ evolution period. 
 
The $^2$H-$^{13}$C $^{\text{RESPIRATION-PLATO}}$CP experiments were performed on a 22.3 T (950 MHz for $^1$H) Bruker NEO  spectrometer using a 2.5 mm HXY triple-resonance probe with 16.67 kHz spinning on uniformly $^{13}$C,$^{15}$N,$^2$H-labeled L-alanine. The pulse sequence in  Fig. \ref{fig:4}a were used with 16 transients, a repetition delay of 0.5 s, and  RF field strengths of 43.5, 39.4, 33.9, and 32.0 for $^2$H RESPIRATION-4 pulses, $^2$H $^{\text{RESPIRATION}}$CP  (4 $\mu$s), $^{\text{RESPIRATION-PLATO}}$CP pulses (4 $\mu$s), $^{13}$C $^{\text{RESPIRATION}}$CP (4 $\mu$s RESPIRATION pulse, 28 $\mu$s $x$, $-x$ pulses) and $^{13}$C $^{\text{RESPIRATION-PLATO}}$CP (peak power; 120 $\mu$s  PLATO sequence). The transfer efficiencies were calculated by integration of the spectral intensities, considering a reference single $\pi/2$ pulse $^{13}$C spectrum, with RF field strengths of 40.3 and relaxation delays of 240 s.

Numerical simulations presented in Supplementary Materials were  performed using the open-source SIMPSON software \cite{SIMPSON,SIMPSONint} using  tensor orientations established using SIMMOL \cite{SIMMOL}. The PLATO pulse sequence was optimized for a single-crystal with principal axis to laboratory frame crystallite orientations of $\beta_{PL}$ = 64.9$^\circ$ by non-linear optimization as described in Ref. \cite{PLATO_adv}.







\clearpage 

%
\bibliography{PLATO_MAS} 
\bibliographystyle{sciencemag}

%
%
%
%
%
%


\section*{Acknowledgments}
Solid-state NMR experiments were carried out at the Danish Center for Ultrahigh-Field NMR Spectroscopy.

\paragraph*{Funding:}
We acknowledge  support from the Aarhus University Research Foundation (AUFF, grant AUFF-E-2021-9-22), the Novo Nordisk Foundation (NERD grant NNF22OC0076002), the Villum Foundation Synergy program (grant 50099), the Danish Independent Research Council (grant 2032-00215B), the Swiss National Science Foundation (Postdoc.Mobility grant 206623), and the DeiC National HPC (g.a. DeiC-AU-L5-0019). We acknowledge the use of NMR facilities at the Danish Center for Ultrahigh-Field NMR Spectroscopy funded by the Danish Ministry of Higher Education and Science (AU-2010-612-181) and the Novo Nordisk Foundation Research Infrastructure - Large Equipment and Facilities program (NNF220C0075797).

\paragraph*{Author contributions:}
All authors contributed to development and discussion of the theory, concept of single-crystal to powder optimization, and translation of static DNP experiments to powder MAS NMR experiments. J.P.C. and A.B.N. performed experiments supported by N.C.N. 
All contributed to writing the manuscript.

\paragraph*{Competing interests:}
There are no competing interests to declare.

\paragraph*{Data and materials availability:}





All data needed to
evaluate the conclusions in the paper are present in the
paper and/or the Supplementary Materials. Raw data, processing, and simulation scripts will be deposited on Zenodo upon
acceptance of the manuscript.




\subsection*{Supplementary materials}
\sloppy
Supplementary Text\\
Figs. S1 - S3\\
References \textit{(38-\arabic{enumiv})}\\ 


\newpage


\renewcommand{\thefigure}{S\arabic{figure}}
\renewcommand{\thetable}{S\arabic{table}}
\renewcommand{\theequation}{S\arabic{equation}}
\renewcommand{\thepage}{S\arabic{page}}
\setcounter{figure}{0}
\setcounter{table}{0}
\setcounter{equation}{0}
\setcounter{page}{1} 


\begin{center}
\section*{Supplementary Materials for\\ \scititle}


José P. Carvalho et al.
 \\
	\small$^\ast$Corresponding author. Email: ncn@chem.au.dk\\

\end{center}

\subsubsection*{This PDF file includes:}
\sloppy
Supplementary Text\\
Figures S1 to S3\\

\newpage






\subsection*{Supplementary Text}

\subsubsection*{Detailed theory}

To demonstrate the  key findings of this paper, namely design of powder MAS dipolar recoupling solid-state NMR experiments from single-crystal optimized static-sample DNP experiments, we will in this section present the string of formula needed to understand the target function used for design of DNP experiments, followed by a proof that this trivially allow for optimization of powder experiments from just a single crystal. Finally, we will justify that the same principles applies translating single-crystal static-sample solid-state NMR experiments with MAS solid-state NMR recoupling on powder samples.

The Hamiltonian governing an electron (S) - nuclear (I) spin-pair subject to an external magnetic field and pulsed microwave (MW) irradiation may to first-order in the rotating frame of the electron spins be written as
\begin{equation}
  \mathcal{H}(t)=\omega_I  I_z+\Delta\omega_S  S_z+A S_z I_z+B S_z I_x+\mathcal{H}_\text{MW}(t) ,
  \label{EqS1}
\end{equation}  
where $\Delta\omega_\text{S}=\omega_\text{S}-\omega_\text{MW}$ denotes the MW offset frequency, $\omega_S$ and $\omega_I$ the electron and nuclear  Larmor frequencies and  
$S_i$ and $I_i$ ($i$ $\in$ $x,y,z$) operators for the electron and nuclear spins.  $A=\boldsymbol{A}_{zz}$ and $B=\sqrt{\boldsymbol{A}^2_{zx}+\boldsymbol{A}^2_{zy}}$  denote amplitudes for the secular and pseudo secular hyperfine coupling with $\boldsymbol{A}$ being the hyperfine coupling tensor. $\mathcal{H}_\text{MW}(t)$ describes the MW pulse sequence with irradiation at the MW carrier frequency $\omega_{MW}$. All frequencies are given in angular frequency units.

Using single-vector effective Hamiltonian theory (SSV-EHT) \cite{SVEHT_EEHT}, the Hamiltonian in the interaction frame of the MW pulse sequence may be written as a Fourier series \cite{BEAM}
\begin{equation}
 \tilde{\mathcal{H}}(t) =  \sum_{\kappa=x,y,z}  \sum_{k =-\infty}^{\infty}  a_{\kappa z}^{(k)} e^{i k \omega_m t} \tilde{S}_\kappa [A I_z + \frac{B}{2} (e^{i k_I \omega_m t} (I_x+iI_y) +e^{-i k_I \omega_m t} (I_x-iI_y))]  + \omega^{(I)}_\text{eff}  I_z-\omega^{(S)}_\text{eff}  \tilde{S}_z ,
  \label{EqS2}
\end{equation}  
with coefficients $a_{\kappa z}^{(k)}$, pulse sequence modulation frequency $\omega_m$,
and effective fields of $\omega_\text{eff}^{(S)}$ and $\omega_\text{eff}^{(I)}=\omega_I-k_I\omega_m$ with $k_I=\text{round}(\omega_I/\omega_m)$ for the electron and nuclear spins, respectively. The pulse sequence modulation frequency simply relates to the duration of the pulse sequence element $\tau_m$ as $\omega_m=2 \pi/\tau_m$. This Hamiltonian provides immediate insight into the design of DNP experiments. Restricting to the weak coupling regime (i.e., $|A|, |B| << |\omega_I|$) and using MW irradiation alone, it is clear that in proximity to a resonance condition $|\omega^{(I)}_\text{eff}| \approx \pm |\omega^{(S)}_\text{eff}|$ and with the effective fields exceeding the effective hyperfine coupling $|\omega^{(I,S)}_\text{eff}| \gtrapprox \frac{B}{2}a_{\kappa z}^{k_I}$, the effective fields along $I_z$ and $\tilde{S}_z$ (see Fig. \ref{fig:1}a) average terms with only one transverse operator. By further noting that the $\tilde{S}_zI_z$ operator can not contribute polarization transfer without RF irradiation on the $I$ spins, only 4 out of 9 operators marked with "+" in Fig. \ref{fig:1}b may be relevant for the design of DNP experiments. This leaves, to first order, the Hamiltonian
\begin{eqnarray}
\overline{\tilde{\mathcal{H}}}^{(1)}&=&\frac{B}{2} [
(a_{xz}^{(-k_I)}+a_{xz}^{(k_I)}) \tilde{S}_x I_x +
i(a_{xz}^{(-k_I)}-a_{xz}^{(k_I)}) \tilde{S}_x I_y  \nonumber \\  &+&
(a_{yz}^{(-k_I)}+a_{yz}^{(k_I)}) \tilde{S}_y I_x + 
i(a_{yz}^{(-k_I)}-a_{yz}^{(k_I)}) \tilde{S}_y I_y
] 
+ \omega^{(I)}_\text{eff}  I_z-\omega^{(S)}_\text{eff}  \tilde{S}_z .
  \label{EqS2b}
\end{eqnarray}

This Hamiltonian may be interpreted as a sum of zero- (ZQ) or double-quantum (DQ) type operators
\begin{equation}
  \overline{\tilde{\mathcal{H}}}^{(1)}=\underbrace{\frac{B}{2} [a^{-} I_x^{23} +b^{-}  I_y^{23}] + \Delta\omega_-^\text{eff}  I_z^{23}]}_\text{ZQ} 
  + \underbrace{\frac{B}{2} [a^{+} I_x^{14} +b^{+}  I_y^{14}] + \Delta\omega_+^\text{eff}  I_z^{14}]}_\text{DQ}  ,
  \label{EqS3a}
\end{equation}
which may conveniently be separated (as resonance should ideally be found hitting only one of them) and individually formulated as
\begin{equation}
  \overline{\tilde{\mathcal{H}}}^{(1)}_{\substack{\scriptscriptstyle{ZQ}\\ \scriptscriptstyle{(DQ)}}}=\frac{B}{2} [a^{\varmp} I_x^{\substack{\scriptscriptstyle{23}\\ \scriptscriptstyle{(14)}}} +b^{\varmp}  I_y^{\substack{\scriptscriptstyle{23}\\ \scriptscriptstyle{(14)}}}] + \Delta\omega_{\varmp}^\text{eff}  I_z^{\substack{\scriptscriptstyle{23}\\ \scriptscriptstyle{(14)}}} ,
  \label{EqS3}
\end{equation}
where we introduced fictitious spin-1/2 operators  \cite{fictitious_vega,fictitious_wokaun_ernst} $I_x^{\substack{\scriptscriptstyle{23}\\ \scriptscriptstyle{(14)}}}=\tilde{S}_x I_x \varpm  
\tilde{S}_y I_y$, 
$I_y^{\substack{\scriptscriptstyle{23}\\ \scriptscriptstyle{(14)}}}=\tilde{S}_y I_x \varmp \tilde{S}_x I_y$, $I_z^{\substack{\scriptscriptstyle{23}\\ \scriptscriptstyle{(14)}}}=\frac{1}{2}(\tilde{S}_z \varmp I_z)$ ,
and the coefficients $\Delta\omega_{\varmp}^\text{eff}=-(\omega^{(I)}_\text{eff} \varpm                       \omega^{(S)}_\text{eff})$, $2a^{\varmp}=(a_{xz}^{(-k_I)}+a_{xz}^{(k_I)}) 
\varpm i (a_{yz}^{(-k_I)}-a_{yz}^{(k_I)})$, and 
$2b^{\varmp}=(a_{yz}^{(-k_I)}-a_{yz}^{(k_I)}) \varmp  i (a_{xz}^{(-k_I)}+a_{xz}^{(k_I)})$. 
The longitudinal linear fields in Eq. (\ref{EqS3a}) allow for selection of either ZQ or DQ operation using pulse sequences with either 
$|\Delta\omega_-^\text{eff}| < |B|$ and 
$|\Delta\omega_+^\text{eff}| >> |B|$ or 
$|\Delta\omega_+^\text{eff}| < |B|$ and 
$|\Delta\omega_-^\text{eff}| >> |B|$, respectively. The large linear field component  truncates the DQ Hamiltonian in the first case, and the ZQ Hamiltonian in the last case.

 By noting that the two phase components in Eq. (\ref{EqS3}) may be described by an effective field $\frac{B}{2}$ scaled by $a_{\varmp}^\text{eff}=\sqrt{(a^{\varmp})^2+(b^{\varmp})^2}$ along $I_x^{\substack{\scriptscriptstyle{23}\\ \scriptscriptstyle{(14)}}}$, an angle (azimuth) $\theta_{\varmp}=\arctan(b^{\varmp}/a^{\varmp})$ describing the rotation around $I_z^{\substack{\scriptscriptstyle{23}\\ \scriptscriptstyle{(14)}}}$, and that the remaining offset term tilts the effective field away from the transverse plane by an angle (zenith) $\phi_{\varmp}=\arctan(\frac{B}{2}a_{\varmp}^\text{eff}/\Delta\omega_{\varmp}^\text{eff})$ to give an effective field  $\omega_{\varmp}=\sqrt{(\frac{B}{2}a_{\varmp}^\text{eff})^2+(\Delta\omega_{\varmp}^\text{eff})^2}$, the first-order effective Hamiltonian may be expressed as
\begin{equation}
  \overline{\tilde{\mathcal{H}}}_{\substack{\scriptscriptstyle{ZQ}\\ \scriptscriptstyle{(DQ)}}}^{(1)}= \omega_{\varmp} \left[
e^{-i \theta_{\varmp} I_z^{\substack{\scriptscriptstyle{23}\\ \scriptscriptstyle{(14)}}}} \left( e^{-i \phi_{\varmp} I_y^{\substack{\scriptscriptstyle{23}\\ \scriptscriptstyle{(14)}}}} I_x^{\substack{\scriptscriptstyle{23}\\ \scriptscriptstyle{(14)}}} e^{i \phi_{\varmp} I_y^{\substack{\scriptscriptstyle{23}\\ \scriptscriptstyle{(14)}}}} \right) e^{-i \theta_{\varmp} I_z^{\substack{\scriptscriptstyle{23}\\ \scriptscriptstyle{(14)}}}} \right] ,
  \label{EqS4}
\end{equation}
with the effective fields and angles illustrated in Fig. ~\ref{fig:1}c. 

DNP transfer from electron spin polarization $\tilde{S}_z=I_z^{14}-I_z^{23}$ to nuclear spin polarization $I_z=I_z^{14}+I_z^{23}$ may be obtained by a ZQ operation inverting $I_z^{23}$ (with $I_z^{14}$  invariant) leading to net positive transfer, or correspondingly by a
DQ operation  inverting $I_z^{14}$ (with $I_z^{23}$ invariant) leading to net negative transfer.  The transfer efficiency may readily be evaluated as 
\begin{eqnarray}
  \langle  I_z\rangle_{\substack{\scriptscriptstyle{ZQ}\\ \scriptscriptstyle{(DQ)}}} (t) &=& 
  \langle  \rho(0)| \tilde S_z\rangle 
  \text{Tr}\left\{  U_{\substack{\scriptscriptstyle{ZQ}\\ \scriptscriptstyle{(DQ)}}}(t) \tilde S_z   U_{\substack{\scriptscriptstyle{ZQ}\\ \scriptscriptstyle{(DQ)}}}^\dagger(t) I_z \right\}
  \nonumber \\
  &=& 
  {\varpm}\langle  \rho(0)| \tilde S_z\rangle \cos^2(\phi_{\varmp})\sin^2(\frac{1}{2}\omega_{\varmp} t)
  \label{EqS5}
\end{eqnarray}
using  $U_{\substack{\scriptscriptstyle{ZQ}\\ \scriptscriptstyle{(DQ)}}}(t)=e^{-i\overline{\tilde{\mathcal{H}}}_{\substack{\scriptscriptstyle{ZQ}\\ \scriptscriptstyle{(DQ)}}}^{(1)} t } = 
e^{-i \theta_{\varmp} I_z^{\substack{\scriptscriptstyle{23}\\ \scriptscriptstyle{(14)}}}} e^{-i \phi_{\varmp} I_y^{\substack{\scriptscriptstyle{23}\\ \scriptscriptstyle{(14)}}}} e^{-i \omega_{\varmp} t I_x^{\substack{\scriptscriptstyle{23}\\ \scriptscriptstyle{(14)}}}} e^{i \phi_{\varmp} I_y^{\substack{\scriptscriptstyle{23}\\ \scriptscriptstyle{(14)}}}} e^{-i \theta_{\varmp} I_z^{\substack{\scriptscriptstyle{23}\\ \scriptscriptstyle{(14)}}}} $ propagated step by step from right to left. Introducing the effective scaled hyperfine coupling constant $B^\text{eff}=\frac{B}{2} a_{\varmp}^\text{eff} $, and writing out the expression in Eq. (\ref{EqS5}) one arrives at
\begin{equation}
  \langle  I_z\rangle_{\substack{\scriptscriptstyle{ZQ}\\ \scriptscriptstyle{(DQ)}}} (t) =
 {\substack{\scriptscriptstyle{+}\\ \scriptscriptstyle{(-)}}} \langle  \rho(0)| \tilde S_z\rangle \frac{(B^\text{eff})^2}{(B^\text{eff})^2+(\Delta\omega_{\varmp}^\text{eff})^2} \sin^2\left[\frac{t}{2} \sqrt{(B^\text{eff})^2+(\Delta\omega_{\varmp}^\text{eff})^2}\right] 
  ,
  \label{EqS6}
\end{equation}
reducing to 
\begin{equation}
  \langle  I_z\rangle_{\substack{\scriptscriptstyle{ZQ}\\ \scriptscriptstyle{(DQ)}}}(t) =
 {\substack{\scriptscriptstyle{+}\\ \scriptscriptstyle{(-)}}} \langle  \rho(0)| \tilde S_z\rangle \sin^2\left(\frac{B^\text{eff}}{2} t\right) 
  \label{EqS7}
\end{equation}
upon matching resonance.

To compute the detailed aspects of orientational averaging in dipolar recoupling experiments, we introduce orientational dependency of $B= T \sqrt{6} d^{2}_{0,1}(\beta_{PL}) = \frac{3}{2} T \sin (2\beta_{PL})$ of the pseudo-secular hyperfine coupling into Eq. (\ref{EqS7}), leading to
\begin{align}
\langle  I_z\rangle_{\substack{\scriptscriptstyle{ZQ}\\ \scriptscriptstyle{(DQ)}}}&={\substack{\scriptscriptstyle{+}\\ \scriptscriptstyle{(-)}}} \langle  \rho(0)| \tilde S_z\rangle \sin^2\left(\frac{3}{8}\, T a_{\varmp}^\text{eff} t  \sin(2\beta_{PL})\right)\\
&={\substack{\scriptscriptstyle{+}\\ \scriptscriptstyle{(-)}}} \langle  \rho(0)| \tilde S_z\rangle \left\{\frac12-\frac12\cos\left[\frac34T a_{\varmp}^\text{eff} t \sin\left(2\beta_{PL}\right)\right]\right\}. 
\end{align} The orientational averaged polarization transfer become
\begin{equation}
\frac1N\int_V\langle I_z\rangle_{\substack{\scriptscriptstyle{ZQ}\\\scriptscriptstyle{(DQ)}}} (t) d\Omega=\varpm\frac12 \langle  \rho_0| \tilde S_z\rangle \int_0^\pi\left\{\frac12-\frac12\cos\left[\frac34T a_{\varmp}^\text{eff} t \sin(2\beta_{PL})\right]\right\}\sin(\beta_{PL})d\beta_{PL} ,\label{eq:S13a}
\end{equation}
which after some algebraic manipulation may be rewritten as
\begin{equation}
\frac1N\int_V\langle I_z\rangle_{\substack{\scriptscriptstyle{ZQ}\\\scriptscriptstyle{(DQ)}}} (t) d\Omega=\varpm\frac12 \langle  \rho_0| \tilde S_z\rangle\left\{1 -\frac12\int_0^\pi \cos\left[\frac34T a_{\varmp}^\text{eff} t \sin(2\beta_{PL})\right]\right\}\sin(\beta_{PL})d\beta_{PL}. \label{eq:s13}
\end{equation}
Defining $z=\frac34T a_{\varmp}^\text{eff} t $ and $\theta=2\beta_{PL}$, we can employ the Jacobi-Anger expansion \cite{abramowitz1968handbook} 
\begin{equation}
\cos(z \sin\theta)=J_0(z)+2\sum_{k=1}^\infty J_{2k}(z)\cos(2k\theta),
\end{equation}
where $J_j$ correspond to  Bessel functions of the first kind. This allows for the integration of a nested trigonometric function, yielding
\begin{equation}
\frac1N\int_V\langle I_z\rangle_{\substack{\scriptscriptstyle{ZQ}\\\scriptscriptstyle{(DQ)}}} (t) d\Omega =
 \varpm \frac12\langle  \rho_0| \tilde S_z\rangle\left[1+J_0\left(\frac34 \xi \right)+\sum_{j=1}^N \frac{2 }{16j^2-1} J_{2j}\left(\frac34 \xi \right)\right]. \label{eq:s15}
\end{equation}

\begin{figure}[H] 
	\centering
	\includegraphics[]{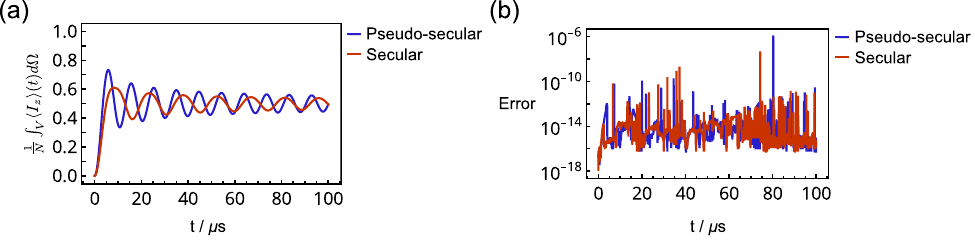} 
	\caption{\textbf{Numerical benchmark of orientational averaging via Bessel functions.} (a) Orientationally averaged nuclear polarization buildup for a pulse sequence with $a_{-}=1.0$, $a_{+}=0.0$, $T/(2\pi)=0.8676$ and $\langle  \rho_0| \tilde S_z\rangle=1.0$. The buildups were computed using Eqs. (\ref{eq:s15}) and (\ref{eq:s22}), considering a pseudo-secular and secular coupling, respectively, and truncating the series at $N=50$. (b) Absolute difference between Eqs. (\ref{eq:s13}) and (\ref{eq:s19}), computed via numerical integration in Mathematica \cite{Mathematica}, and Eqs. (\ref{eq:s15}) and (\ref{eq:s22}), truncated at $j=50$.}
	\label{fig:S1} %
\end{figure}

To expand the scope of this formulation to cover single-crystal optimization for dipolar recoupling involving secular terms (as encountered for the term proportional to $A$ in Eq. (\ref{EqS1}) for DNP supplemented with RF irradiation on the nuclear spins or for the secular dipolar coupling for dipolar recoupling in solid-state NMR), we also consider a  coupling with orientational dependence $A=2 T d^2_{0,0}(\beta_{PL}) = T (3 \cos^2 (\beta_{PL})-1)$.  We note that for solid-state NMR recoupling applications this term has to be halved as the spin operators is typically formulated as $2S_zI_z$ rather than $S_zI_z$ typically used in EPR, and also in this paper starting from DNP. In this case Eq. (\ref{EqS7}) translates to  
\begin{equation}
    \langle  I_z\rangle_{\substack{\scriptscriptstyle{ZQ}\\ \scriptscriptstyle{(DQ)}}}=\varpm\frac12 \langle  \rho_0| \tilde S_z\rangle  \sin^2\left[\frac{1}{4}\, T a_{\varmp}^\text{eff} t  \left(3\cos^2({\beta_{PL}})-1\right)\right]
\end{equation}
which is simplified to
\begin{equation}
\langle I_z\rangle_{\substack{\scriptscriptstyle{ZQ}\\ \scriptscriptstyle{(DQ)}}}=\varpm\frac12 \langle  \rho_0| \tilde S_z\rangle  \left(\frac12-\frac12\cos\left\{\frac14T a_{\varmp}^\text{eff} t \left[1+3\cos\left(2\beta_{PL}\right)\right]\right\}\right).
\end{equation}
The orientational averaged polarization transfer is then given by
\begin{equation}
\frac1N\int_V\langle I_z\rangle_{\substack{\scriptscriptstyle{ZQ}\\\scriptscriptstyle{(DQ)}}} (t) d\Omega=\varpm\frac12 \langle  \rho_0| \tilde S_z\rangle  \int_0^\pi\left\{\frac12-\frac12\cos\left[\frac14T a_{\varmp}^\text{eff} t (1+3\cos(2\beta_{PL}))\right]\right\}\sin(\beta_{PL})d\beta_{PL} \label{eq:S13b},
\end{equation}
which can be expanded as
\begin{multline}
\frac1N\int_V\langle I_z\rangle_{\substack{\scriptscriptstyle{ZQ}\\\scriptscriptstyle{(DQ)}}} (t) d\Omega=\varpm\frac12 \langle  \rho_0| \tilde S_z\rangle  \left(1-\frac12 \int_0^\pi\left\{\cos\left(\frac14 T a_{\varmp}^\text{eff} t \right)\cos\left[\frac34 T a_{\varmp}^\text{eff} t \cos\left(2\beta_{PL}\right)\right]\right.\right.\\\left.\left.-\sin\left(\frac14 T a_{\varmp}^\text{eff} t \right)\sin\left[\frac34 T a_{\varmp}^\text{eff} t \cos\left(2\beta_{PL}\right)\right]\right\}\sin(\beta_{PL})d\beta_{PL} \right) \label{eq:s19}
\end{multline}
Once more, we define  $z=\frac34T a_{\varmp}^\text{eff} t $ and $\theta=2\beta_{PL}$ and employ the Jacobi-Anger identities \cite{abramowitz1968handbook} 
\begin{align}
\sin(z \sin\theta)=2\sum_{k=0}^\infty J_{2k+1}(z)\sin[2(k+1)\theta],\\
\cos(z \cos\theta)=J_0(z)+2\sum_{k=1}^\infty(-1)^k J_{2k}(z)\cos(2k\theta),
\end{align}
to allow for the integration of the trigonometric functions. This yields
\begin{multline}
\frac1N\int_V\langle I_z\rangle_{\substack{\scriptscriptstyle{ZQ}\\\scriptscriptstyle{(DQ)}}} (t) d\Omega=\varpm\frac12\langle  \rho_0| \tilde S_z\rangle\left[1-J_0\left(\frac{3T a_{\varmp}^\text{eff} t }{4}\right)\cos\left(\frac{T a_{\varmp}^\text{eff} t }{4}\right)\right.+\\\left.\sum_{j=1}^\infty \frac{2(-1)^j}{16j^2-1}J_{2j}\left(\frac{3T a_{\varmp}^\text{eff} t }{4}\right)\cos\left(\frac{T a_{\varmp}^\text{eff} t }4\right)+\sum_{j=1}^\infty \frac{2(-1)^j}{16j^2-16j+3}J_{2k-1}\left(\frac{3T a_{\varmp}^\text{eff} t }{4}\right)\sin\left(\frac{T a_{\varmp}^\text{eff} t }4\right)\right] \label{eq:s22}
\end{multline}
In Fig. \ref{fig:S1}, Eqs. (\ref{eq:s15}) and (\ref{eq:s22}) were benchmarked against Eqs. (\ref{eq:s13}) and (\ref{eq:s19}), computed via numerical integration, demonstrating their validity.

\subsection*{Supplementary numerical simulations}

Supplementing  theoretical and experimental descriptions in the main tetext,e present here numerical simulations of the broadband performance of the $^{\text{PLATO}}$CP and 
$^{\text{RESPIRATION-PLATO}}$CP obtained using the open-source SIMPSON software \cite{SIMPSON,SIMPSONint}. 

\begin{figure} 
	\centering
	\includegraphics[width=0.8\textwidth]{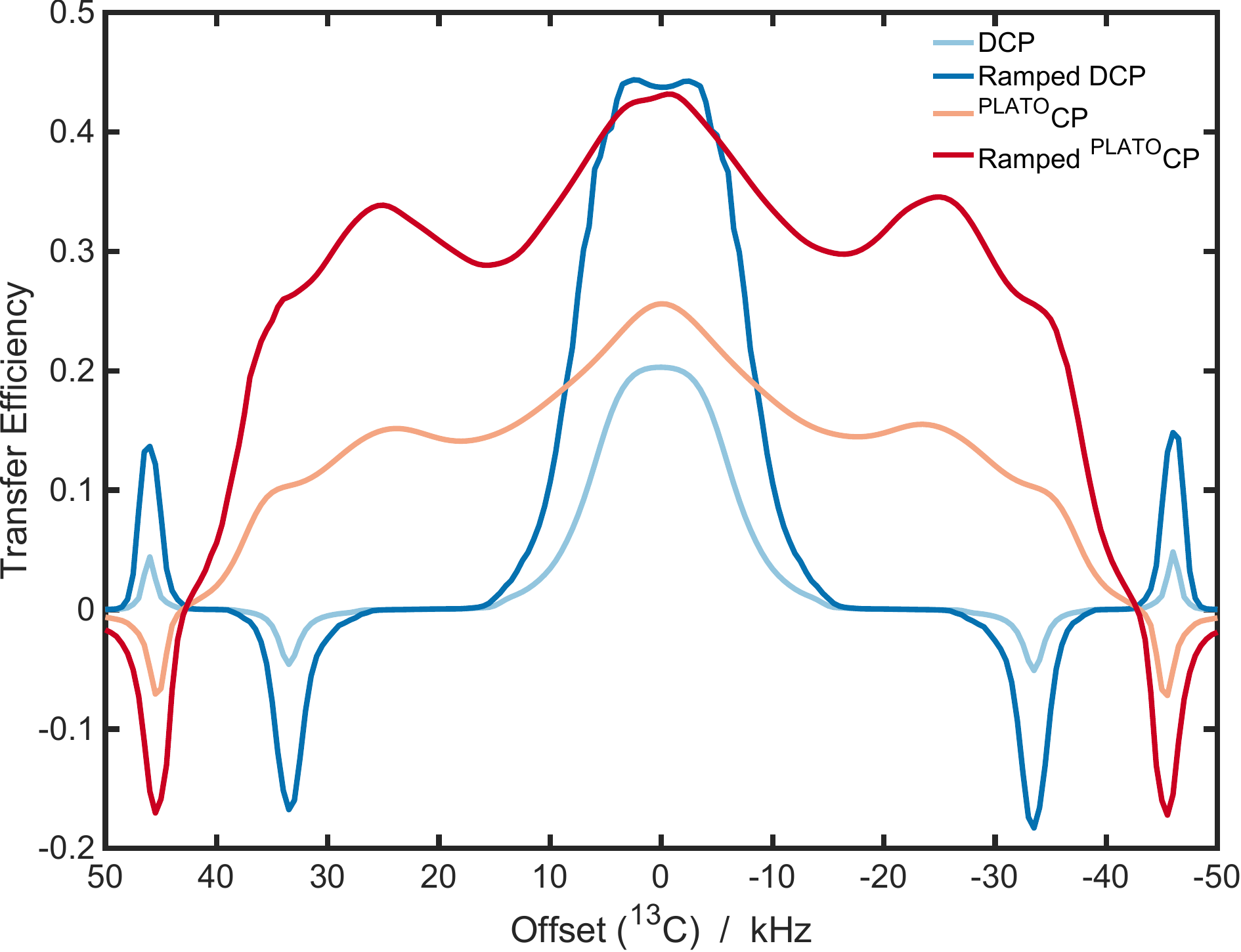} 
	\caption{\textbf{Numerical simulations of $^{15}$N $\rightarrow$ $^{13}$C CP MAS experiments.} Numerical simulation of DCP, ramped DCP, $^{\text{PLATO}}$CP, and ramped $^{\text{PLATO}}$CP experiments for a $^{15}$N-$^{13}$C spin-pair under conditions of a static magnetic field of 16.4 T (700 MHz for $^1$H) and 25.000 Hz MAS. Parameters given in Supplementary Text.}
	\label{fig:S2} 
\end{figure}

Figure \ref{fig:S2} compares the offset profiles from double-cross-polarization 
(DCP), ramped (adiabatic) DCP (70-100\% linear ramp on $^{15}$N), $^{\text{PLATO}}$CP, and ramped (adiabatic) $^{\text{PLATO}}$CP (90-100\% linear ramp on $^{15}$N) 
under conditions matching the experiments in Fig. \ref{fig:3} for a representative
$^{15}$N-$^{13}$C$_\alpha$ spin-pair system. The simulations were calculated for the NC pulse sequence elements shown in Fig. \ref{fig:3}a (bottom) with the initial density operator being $x$-phase coherence on the $^{15}$N spin ($I_x$) and detection of $x$-phase coherence on the $^{13}$C spin ($S_x$). The simulations assumed a spin rate of $\omega_r/(2\pi)$ = 25.000 Hz, an external magnetic field of 16.4 T (corresponding to 700 MHz for $^1$H). The DCP simulation assumed 5.000 and 30.000 Hz RF field strength on $^{15}$N and $^{13}$C, respectively, and a mixing time of 1.28 ms (32 rotor periods). Ramped DCP assumed 30.000 Hz RF field strenght on $^{13}$C and a 70-100\% RAMP on $^{15}$N extending from 4.100 to 5.900 Hz over a mixing time of 7 ms. $^{\text{PLATO}}$CP used a constant $^{15}$N RF field of amplitude 10.224 Hz and a PLATO sequence on $^{13}$C of length 120 $\mu$s (see amplitudes in Materials and Methods) repeated 16 times to give a mixing time of 1.92 ms. The ramped $^{\text{PLATO}}$CP simulation used PLATO irradiation on the $^{13}$C RF channel repeated 55 times to give a mixing time of 6.6 ms, and a 90-100\% ramp extending from 9.725 to 10.725 Hz on $^{15}$N for this period. The simulations assume powder averaging using 20 ($\alpha_{PR}$, $\beta_{PL}$) uniformly distributed angles obtained by the REPULSION method \cite{REPULSION} and 5 $\gamma_{PR}$ angles incorporated using $\gamma$-COMPUTE \cite{g-COMPUTE}. Tensor orientations were established using SIMMOL \cite{SIMMOL}. The spin system parameters used for the simulations were 
$\omega_\delta^{iso}/(2\pi)$($^{15}$N) = 0 ppm, $\omega_\delta^{aniso}/(2\pi)$($^{15}$N) = 5 ppm, $\eta_\delta$($^{15}$N) = 0.00, and 
$\{\alpha_{PR},\beta_{PL},\gamma_{PR}\}_\delta$($^{15}$N) = 
$\{-90^\circ,-90^\circ,-17^\circ\}$,
$\omega_\delta^{iso}/(2\pi)$($^{13}$C) = 0 ppm, $\omega_\delta^{aniso}/(2\pi)$($^{13}$C) = -20 ppm, $\eta_\delta$($^{13}$C) = 0.43, and 
$\{\alpha_{PR},\beta_{PL},\gamma_{PR}\}_\delta$($^{13}$C) = 
$\{90^\circ,90^\circ,0^\circ\}$,
b$_{NC}/(2\pi)$ = 988 Hz, 
$\{\alpha_{PR},\beta_{PL},\gamma_{PR}\}_{NC})$ = 
$\{0^\circ,90^\circ,115^\circ\}$, and $J_{NC}$ = 11 Hz.

\begin{figure} 
	\centering
	\includegraphics[width=0.8\textwidth]{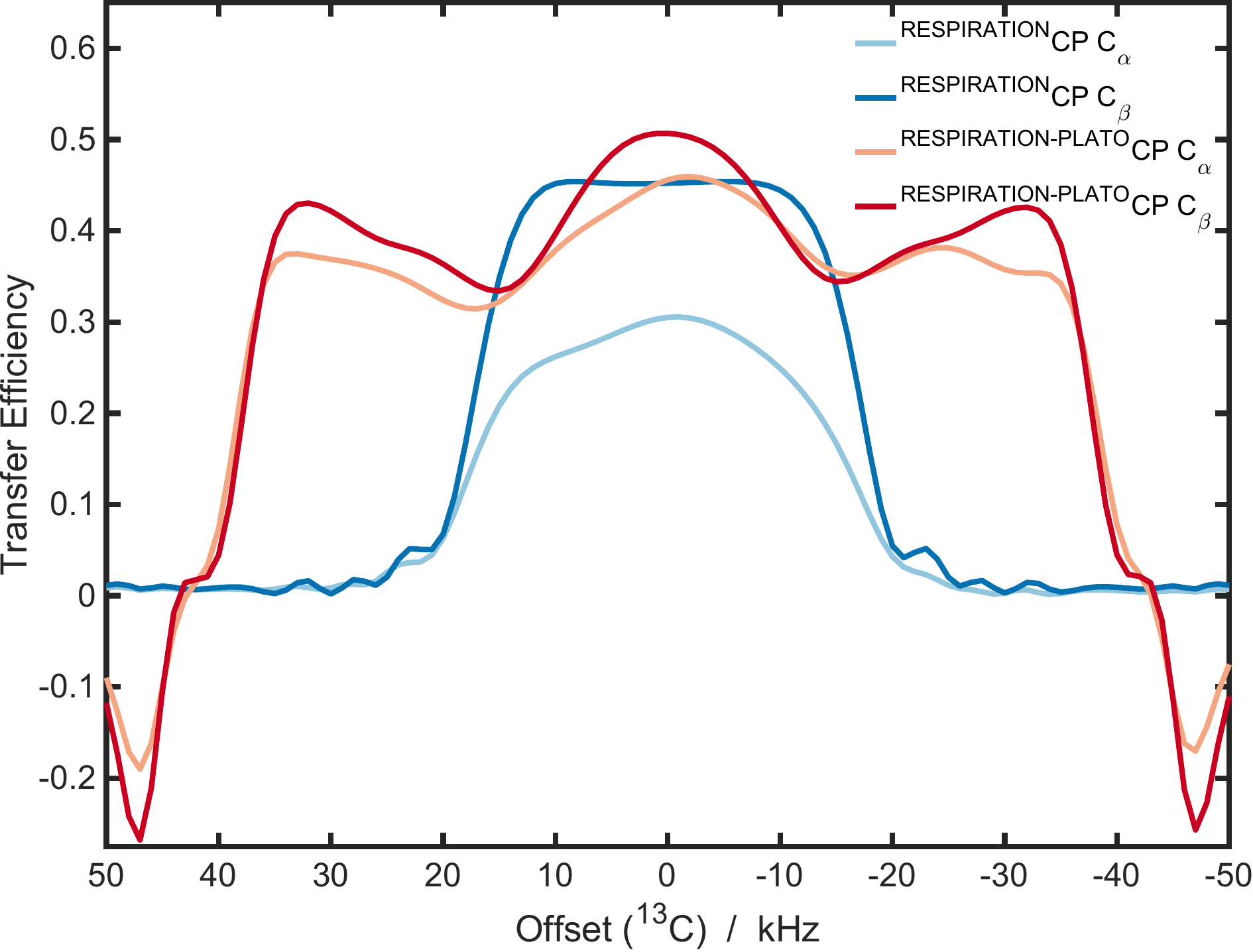} 
	\caption{\textbf{Numerical simulations of $^{2}$H $\rightarrow$ $^{13}$C CP MAS experiments.}  Numerical simulation of $^{13}$C$_\alpha$ and $^{13}$C$_\beta$ offset profiles for $^2$H $\rightarrow$ $^{13}$C $^{\text{RESPIRATION}}$CP and $^{\text{RESPIRATION-PLATO}}$CP experiments under conditions of  22.3 T (950 MHz for $^1$H) and 16.667 Hz MAS. See parameters in Supplementary Text.}
	\label{fig:S3} 
\end{figure}

Figure \ref{fig:S3} shows numerical offset profiles from $^{\text{RESPIRATION}}$CP and $^{\text{RESPIRATION-PLATO}}$CP experiments under conditions matching the experiments in Fig. \ref{fig:4} calculated for representative
 $^{2}$H-$^{13}$C$_\alpha$ and $^{2}$H-$^{13}$C$_\beta$ spin-pair systems. The simulations included an initial RESPIRATION-4 pulse \cite{OC_2H} (as shown in Fig. \ref{fig:4}a) with the initial density operator being polarization on the $^{2}$H spin ($I_z$) and detection of $x$-phase coherence on the $^{13}$C spin ($S_x$). The simulations assumed a spin rate of $\omega_r/(2\pi)$ = 16.667 Hz, an external magnetic field of 22.3 T (corresponding to 950 MHz for $^1$H). All simulations were initialized with a RESPIRATION-4 pulse \cite{OC_2H} on $^2$H using an RF field strength of 43.000 Hz, with each of the 4 pulses having a duration of 1.45 $\mu$s, separated by a rotor period (60 $\mu$s). The $^{\text{RESPIRATION}}$CP and $^{\text{RESPIRATION-PLATO}}$CP used a 4 $\mu$s pulse of amplitude 34.000 Hz for each rotor period during a total mixing time of 10 rotor periods (0.6 ms), corresponding to 10 RESPIRATION-CP elements and 5 $^{\text{RESPIRATION-PLATO}}$CP elements. The $^{\text{RESPIRATION}}$CP experiment used an RF amplitude of 33.333 Hz (twice the spinning frequency) for the $x$ and $-x$ spin lock pulses, and 31.500 Hz for the 4 $\mu$s RESPIRATION pulses. The $^{\text{RESPIRATION-PLATO}}$CP experiment used the PLATO sequence (see Materials and Methods) extended over two rotor periods on the $^{13}$C RF channel. 
Simulations assume powder averaging using 66 ($\alpha_{PR}$, $\beta_{PL}$) uniformly distributed angles obtained by the REPULSION method \cite{REPULSION} and 8 $\gamma_{PR}$ angles incorporated using $\gamma$-COMPUTE \cite{g-COMPUTE}. Tensor orientations were established using SIMMOL \cite{SIMMOL}. 
 The spin system parameters for the $^2$H-$^{13}$C$_\alpha$ spin-pair were
$\omega_\delta^{iso}/(2\pi)$($^{2}$H) = 1.15 ppm, $\omega_\delta^{aniso}/(2\pi)$($^{2}$H) = 7.7 ppm, $\eta_\delta$($^{2}$H) = 0.65, and 
$\{\alpha_{PR},\beta_{PL},\gamma_{PR}\}_\delta$($^{2}$H) = 
$\{-162^\circ,66^\circ,67^\circ\}$
$C_{Q}/(2\pi)$($^{2}$H) = 150 kHz, $\eta_Q$($^{2}$H) = 0.0, and 
$\{\alpha_{PR},\beta_{PL},\gamma_{PR}\}_Q$($^{2}$H) = 
$\{0^\circ,0^\circ,0^\circ\}$
$\omega_\delta^{iso}/(2\pi)$($^{13}$C) = 16 ppm, $\omega_\delta^{aniso}/(2\pi)$($^{13}$C) = -20 ppm, $\eta$($^{13}$C) = 0.43, and 
$\{\alpha_{PR},\beta_{PL},\gamma_{PR}\}_\delta$($^{13}$C) = 
$\{-109^\circ,97^\circ,39^\circ\}$
b$_{^2HC}/(2\pi)$ = -3689 Hz and 
$\{\alpha_{PR},\beta_{PL},\gamma_{PR}\}_{^2HC})$ = 
$\{0^\circ,14^\circ,-94^\circ\}$, and $J_{^2HC}$ = 21.5 Hz.
For the $^2$H-$^{13}$C$_\beta$ spin-pair, the parameters were
$\omega_\delta^{iso}/(2\pi)$($^{2}$H) = -1.15 ppm, $\omega_\delta^{aniso}/(2\pi)$($^{2}$H) = 7.7 ppm, $\eta_\delta$($^{2}$H) = 0.65, and 
$\{\alpha_{PR},\beta_{PL},\gamma_{PR}\}_\delta$($^{2}$H) = 
$\{-162^\circ,66^\circ,67^\circ\}$
$C_{Q}/(2\pi)$($^{2}$H) = 53 kHz, $\eta_Q$($^{2}$H) = 0.1, and 
$\{\alpha_{PR},\beta_{PL},\gamma_{PR}\}_Q$($^{2}$H) = 
$\{0^\circ,0^\circ,0^\circ\}$
$\omega_\delta^{iso}/(2\pi)$($^{13}$C) = -16 ppm, $\omega_\delta^{aniso}/(2\pi)$($^{13}$C) = 10 ppm, $\eta$($^{13}$C) = 0.0, and 
$\{\alpha_{PR},\beta_{PL},\gamma_{PR}\}_\delta$($^{13}$C) = 
$\{-98^\circ,123^\circ,85^\circ\}$
b$_{^2HC}/(2\pi)$ = -3689 Hz and 
$\{\alpha_{PR},\beta_{PL},\gamma_{PR}\}_{^2HC})$ = 
$\{0^\circ,14^\circ,-94^\circ\}$, $J_{^2HC}$ = 21.5 Hz.

\subsection*{Supplementary Figures}


\end{document}